\documentclass[aps,prb,preprint,showpacs,superscriptaddress,floatfix]{revtex4}
\usepackage{graphics}
\usepackage{amsmath}
\usepackage{amssymb}
\usepackage{calc}
\usepackage{epsfig}
\usepackage{color}
\usepackage{amsmath,bm}

\begin{document}

\title{Coupling of conduction electrons to two-level systems formed by hydrogen:
A scattering approach}

\author{I. Nagy}
\affiliation{Department of Theoretical Physics, Institute of
Physics,\\ Technical University of Budapest, H-1521 Budapest, Hungary}
\affiliation{Donostia International Physics Center DIPC, P. Manuel de Lardizabal 4,\\
20018 San Sebasti\' an, Spain}
\author{A. Zawadowski}
\affiliation{Department of Theoretical Physics, Institute of 
Physics,\\ Technical University of Budapest, H-1521 Budapest, Hungary}
\affiliation{Research Institute for Solid State Physics and Optics,
Hungarian Academy of Sciences,\\ H-1525 Budapest, Hungary}

\date{\today}

\begin{abstract}

An effective Hamiltonian which could model the interaction between a
tunneling proton and the conduction electrons of a metal is
investigated. A remarkably simple correlation between the motion of
the $TLS$-atom and an angular-momentum change of scattering electron
is deduced, at the first-order Born level, by using a momentum-space
representation with plane waves for initial and final states. It is
shown that the angular average of the scattering amplitude-change at
the Fermi surface depends solely on the difference of the first two
phase shifts, for small-distance displacements of the heavy
particle. For such a limit of displacement, and within a
distorted-wave Born approximation for initial and final states, the
change in the scattering amplitude is expressed via trigonometric
functions of scattering phase shifts at the Fermi energy. The
numerical value of this change is analyzed in the framework of a
self-consistent screening description for impurity-embedding in a
paramagnetic electron gas. In order to discuss the so-called
antiabatic limit on the same footing, a comparison with matrix
elements obtained by the potential-gradient of an unscreened Coulomb
field is given as well. The coupling of the tunneling proton to a
free-electron-like electron gas is in the typical range obtained, by
ultrasound experiments for different metallic glasses, from
scattering rates for a Korringa-type relaxation process. That
coupling is too weak to be in the range required for realization of
the two-channel Kondo effect.

\end{abstract}


\maketitle

\section{Introduction}

In the last decades hydrogens in metals deserved very extensive experimental studies
and vast theoretical considerations.
In crystalline solid the hydrogen ($H$) sits in a well-defined
interatomic position like, e.g., in $Pd$ or $Pt$. That is not the case in amorphous systems
and at dislocations and other nonperiodic distortions.
If the $H$ has some more room between the host atoms its position may be not well-defined and
it moves between two positions.
Such systems are known as two-level systems ($TLS$) and they have been very
extensively studied \cite{Beck81}.
The coupling to the conduction electrons can result in an extra contribution
to the electrical resistivity \cite{Black79}.
If the atom has two metastable positions the electron scattering amplitude
in different angular momentum channels
depends on the atomic position.
The difference between these amplitudes is described in the literature
by a coupling $V^z$. That coupling contributes to the resistivity in a conventional way.

There are, however, other couplings where electrons induce
transitions between the two levels. Thus an assisted transition can
be realized by the tunneling of the atom between the two positions
\cite{Cochrane75,Zawadowski80,Vladar83}. The importance of that
coupling, denoted by $V^{x}$ and $V^{y}$, is highly debated
\cite{Aleiner01,Zarand05}. These models consider different atoms
with sizeable differences in their masses. The original suggestion
is limited to small tunneling rate, while more intensive tunneling
induces an essential split between the energies of the atomic
eigenstates which reduces their roles. The model has attracted
considerable interest as it was suggested that at low temperature
exhibits non Fermi liquid behavior known as the two-channel Kondo
($2CK$) effect, which has its own theoretical interest
\cite{Ralph92, Cox98}.

Experimental supports come much more from the studies of point contacts than from direct
measurement of the electrical resistivity \cite{Ralph92}. The Cornell group made
detailed suggestion how the observed zero-bias anomalies could be due to that $2CK$ effect
which has been also highly debated \cite{Aleiner01}. Since that time the original model was modified
\cite{Zarand05}
by taking into account the actual electronic structure at the $TLS$.
The possibility was also considered \cite{Borda03} where the atom moves between the two positions via
the next higher energy level of the atomic motion.
These suggestions were aimed to increase the electron assisted amplitudes ($V^{x}$, $V^{y}$)
to make the $2CK$ more feasible. The enhancement of these couplings by the
renormalization due to the conduction electrons is drived by $V^{z}$. The estimated
couplings were on the borderline, therefore, those should be studied in more details.
It is crucial whether those coupling strength could reach some certain regions
which are very sensitive on the strength of $V^{z}$.

Recently it was suggested that $H$ is a possible candidate
\cite{Ralph92}. Indeed, zero-bias anomalies have been observed in
hydrogenated $Pd$ point contact with considerable size
\cite{Csonka04}. That system could deserve more extensive
experimental studies. Such zero-bias anomalies were also observed in
hydrogenated $Pt$ point contact \cite{Djukic05}. The idea has beeen
also raised that other systems may contain some water with
metastable positions of the hydrogen atoms.

These recent developments justifies further studies of the coupling
of a heavy tunneling particle to conduction electrons. The
satisfactory realiable estimation of these coupling could be
performed by considering certain atomic configuration with detailed
knowledge of the electronic density of states in the region of the
$TLS$. In other words, a material-specific description of the {\it
host} could be based upon more realistic but much more complex
specifications. Thus, a calculation of the coupling would be too
ambitious in case where the structural surrounding would require
more knowledge.

The present paper is devoted to a minimal program. An embedded
proton is considered in a double-well potential which determines the
two possible positions of the heavy particle. These positions are
considered as stable ones. The main task is to take into account the
screening, due to the Coulomb interaction and charge-response, in a
satisfactory way. In this work the ground-state screening in the
surrounding electron gas is described in a self-consistent
Hartree-like manner, by using Kohn-Sham independent single-particle
states whose occupation is prescribed by the Pauli principle. The
change in the scattering amplitude, due to a potential-displacement,
is expressed in our renormalized perturbation theory via
matrix-elements taken between stationary scattering eigenstates of
the external Kohn-Sham field, instead of commonly applied
matrix-elements between unperturbed plane-wave states of a system
without impurity. Concretely, in the present work we shall apply the
distorted wave Born approximation \cite{Ballentine98} for scattering
characteristics.

The paper is organized as follows. The next section, Sec. II, is
devoted to the details of our theory and a discussion of the
plane-wave-based, first-order Born approximation for matrix-elements
is also given. These results, obtained by using screened potentials,
are compared with those based on an unscreened Coulomb field. The
last section, Sec. III, contains a short summary and an outlook for
further possible developments. The Appendix summarizes the commonly
applied basic elements of previous motivating works.

\newpage
\section{Theory and results}

As we motivated in the Introduction, an atom is considered which has two stable
positions in metallic matrix due to
a double potential well. The positions are symmetric to a central point and they
are at $z={\pm}\frac{d}{2}$, respectively. This is the standard picture to a theoretically
active field with considerable experimental relevances;
see Figure 1 for illustration.

\begin{figure}[tbp]
\resizebox{6.5cm}{!}{\includegraphics*{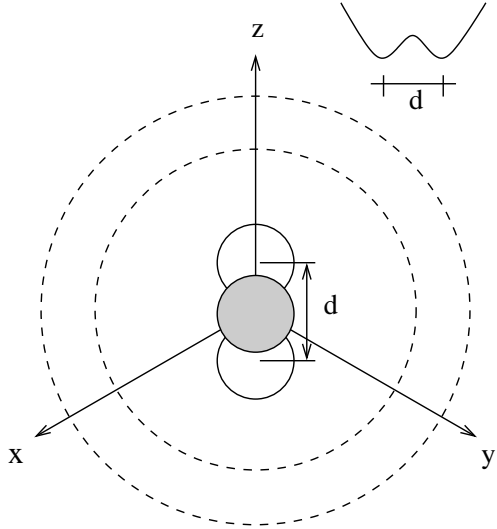}}
\caption{Illustration of the two-level system.
The shadowed circle represents the screened proton in the central position and the two
others, separated by $d$, the TLS. The dashed lines correspond to the spherical wave centered
around the origin. The inset shows the potential of the TLS.}
\label{figure1}
\end{figure}

The main goal of the present contribution is to determine the {\it
change} in the electron scattering amplitude in different
angular-momentum channels when the atom is moved out from the
central point to one of the two positions, but the set of the
electron wave functions is still centered at the origin. These
positions, at $z={\pm}\frac{d}{2}$ as the illustrative Figure shows,
are described by a pseudospin $\sigma^{z}=\pm{1}$.

The atom is embedded in the degenerate electron gas of the metallic target. The general form
of the Hamiltonian is $H=H_0\, +\, H_1$, where the diagonal $H_0$ matrix
stands for independent electrons in stationary eigenstates of a self-consistent external field $V(r)$

\begin{equation}
H_0\,=\, \sum_{{\gamma},\sigma}\, {\varepsilon}_{\gamma}\, a^{\dag}_{{\gamma}\sigma}\, a_{{\gamma}\sigma},
\end{equation}
in which the ${\varepsilon}_{\gamma}$ are energy eigenvalues of bound and scattering eigenstates.
The $a^{\dag}_{{\gamma}\sigma}$ and $a_{{\gamma}\sigma}$ create and annihilate these eigenstates
of Eq.(1) of spin $\sigma$.

In order to show the difference of our distorted-wave Born method
from the commonly applied plane-wave $(pw)$ approximation, and thus
provide a clear phenomenology to understanding, we start by assuming
a weak potential-energy and write $H_1^{(pw)}$ in this
unperturbed-state representation for initial and final states as

\begin{equation}
H_1^{(pw)}\,=\, \sigma^{z}\, \frac{1}{V}\, \sum_{{\bf q}}\,
V^{z}_{pw}({\bf q})\, \sum_{{\bf p}}\, b^{\dag}_{{\bf p}+{\bf q}}\,
b_{{\bf p}},
\end{equation}
with the corresponding operators for creation and annihilation of
plane-wave states. Here we have ${\bf q}={\bf k_2}-{\bf k_1}$, and
thus $V^{z}_{pw}({\bf q})$
is given by

\begin{equation}
V^{z}_{pw}({\bf q})\, =\, \int\, d{\bf r}\, e^{-i{\bf q}\cdot{\bf
r}}\, {\Delta}V({\bf r}).
\end{equation}
Since the perturbation ${\Delta}V({\bf r})$, due to the shift in the
atomic position, is real we have the $[V^{z}_{pw}({\bf
q})]^{*}=V^{z}_{pw}(-{\bf q})$ character. In other words, the
$H_1^{(pw)}$ is Hermitian. The change in the scattering amplitude is
${\Delta}F_{pw}({\bf q})=[m/(2\pi{\hbar}^2)]V^{z}_{pw}({\bf q})$.
The extended continuous system can be recovered simply by using the
prescription $(1/V){\sum}_{\bf q}\rightarrow{\int d{\bf
q}/(2\pi)^3}$. The ${\bf q}$-summation must be performed at the
Fermi level to respect the Pauli's principle, and thus the
determinant-character \cite{Peierls79} of the state vector of $H_0$.
This will result in, by using the $q=2k_F\sin({\vartheta}/2)$ exact
expression for the momentum change in elastic scattering, an
averaging (see below) over the scattering angle. From this point of
view, our procedure is similar to the one \cite{Morel62,Ashcroft04}
used in the theory of conventional superconductivity to define a
repulsive (electron-electron) pseudopotential, from a spherical
$V(q)$ at the Fermi surface.

One can calculate the change [${\Delta}F({\bf k_2},{\bf k_1})$] in
the scattering amplitude beyond the usual Born approximation, by
using expansions for properly defined \cite{Ballentine98} initial
and final states in terms of normalized spherical harmonics.
Furthermore, it was suggested \cite{Zawadowski80} that one can
define a Hermitian $H_1$, beyond the weak-coupling limit, in the
following way

%
\begin{equation}
H_1\,=\, \sigma^{z}\, \frac{1}{V}\, \sum_{{\bf k_1}{\bf k_2}}\,
V^{z}_{{\bf k_2}{\bf k_1}}\, a^{\dag}_{{\bf k_2}\sigma}\, a_{{\bf
k_1}\sigma},
\end{equation}
where ${V}^{z}_{{\bf k_2}{\bf
k_1}}=\sum_{\alpha\beta}f^{*}_{\beta}(\hat{k_2})V^{z}_{\alpha\beta}(k_1,k_2)
f_{\alpha}(\hat{k_1})$ is a {\it suitable} \cite{Zawadowski80}
coupling, mediated now, in our case, by a ${\Delta}V({\bf r})$
perturbation. The intruduced \cite{Vladar83}
$f_{lm}(\hat{k})=i^l\sqrt{4\pi}Y_{lm}(\hat{k})$ functions are
spherical harmonics, and $\alpha$ and ${\beta}$ run over a properly
chosen set (see below) of angular momentum indices $(l,m)$.
The present calculation is carried out in several steps.

$(i)$ {\it Calculations of the screened potential around the proton and the phase shifts.}
The charge is sitting in the central position where its screened field $V(r)$ and a
complete set of one-electron wave functions of occupied states are determined in a self-consistent way.
This is achieved, in practice, by applying the Kohn-Sham method of density-functional theory ($DFT$),
which reduces the complicated many-body problem of the inhomogeneous electron gas (in the presence
of a charge)
to a single particle problem \cite{Kohn99}. The calculations are performed for a grand-canonical
system, i.e., at a fixed chemical potential, applying local-density approximation ($LDA$) for
the exchange-correlation potential. The single-particle potential energy $V(r)$ has a simple form in
this approximation

\begin{equation}
V(r)\, =\, -\, \frac{Z e^2}{r}\, +\, {\int}\, d^{3}{\bf r'}\, \frac{{\Delta}n(r')}{|\bf {r-r'}|}\,
+\, {\Delta}{\nu}_{xc}[n(r)],
\end{equation}
in which ${\Delta}n(r)$ is the screening density.
The many-body term ${\Delta}{\nu}_{xc}[n]$ is expressed via an input exchange-correlation
chemical potential $(\mu_{xc})$ as ${\Delta}{\nu}_{xc}[n]=\mu_{xc}(n_0+{\Delta}n)-\mu_{xc}(n_0)$, in order
to have a vanishing effective potential energy at infinity.

For a given density ($n_0$) of the screening environment, and
depending on the magnitude of the attractive embedded charge, the
total density consists of bound and scattering eigenstates. The
$n(r)=n_0+{\Delta}n(r)$ total density, the basic variable of $DFT$,
is constructed out by summing over doubly occupied bound and
scattering-like states

\begin{equation}
\psi_{\bf k}^{\pm}({\bf r})\, =\, \sqrt{{4\pi}}\, \sum_{lm}\,
A_{l}(k)\, R_l(k,r)\, Y^{*}_{lm}(\hat r)\, f_{lm}({\hat k}),
\end{equation}
in which $R_l(k,r)$ are self-consistent solutions of the radial
Kohn-Sham equations with $V(r)$ at $({\hbar}k)^2/2m$ scattering
energy, and $A_l(k)=e^{{\pm}i{\delta}_l(k)}$ where ${\delta}_l(k)$
is the phase shift. The above continuum states are normalized on the
$k$-scale, and thus the scattering ($sc$) part of the induced
density comes from an integral over the Fermi-Dirac distribution
function

\begin{equation}
{\Delta}n_{sc}(r)\, =\, \frac{1}{\pi^2}\, \sum_{l=0}^{\infty}\, (2l+1)\, \int_{0}^{k_F}\, dk\, k^2\,
\left[R_l^2(k,r) - j_l^2({k}r)\right].
\end{equation}
The total-screening condition, $4\pi\int_{0}^{\infty}drr^2 {\Delta}n(r)=Z$, implies
the Friedel sum rule of scattering phase shifts in one-electron mean-field treatments

\begin{equation}
Z\, =\, \frac{2}{\pi}\, \sum_{l=0}^{\infty}\, (2l+1)\, {\delta}_l(k_F)\,
-\, \frac{2}{\pi}\, \sum_{l=0}^{\infty}\, (2l+1)\, {\delta}_l(0)\, +\, N_b.
\end{equation}
The rule is satisfied, of course, at numerical self-consistency of iterations.
Here $N_b$ denotes the number of occupied bound states.
In a ground-state calculation, on which the present work is based,
the last two terms cancel each other according to Levinson's theorem \cite{Almbladh76}.
At metallic densities, already
the first few phase-shifts
provide a very accurate approximation.
For example, at the $r_s=2.5$ value of the density parameter
one has ${\delta}_0(k_F)=1.2213$ and ${\delta}_1(k_F)=0.0894$
in the case of a proton.
This effective $r_s$ can characterize the mobile part
of the electron fluid of a $Pd$ target, and satisfies \cite{Perrot94} a necessary condition
appropriate for correctly describing ground state properties of defects.

Illustrative results are exhibited in Figure 2 for the leading, $l=0$ and $l=1$,
radial wave functions calculated at the $r_s=2.5$ value of the Wigner-Seitz parameter.
The $l=0$ component shows a Coulomb-like enhancement, $\sqrt{2\pi/k_F}$ at the origin $r=0$,
over the plane-wave-based unity.
The deviations from the plane-wave components $j_0(k_Fr)$ and $j_1(k_Fr)$ are notable,
as Figure 3 shows via the corresponding products of components.
The $R_0(k_F,r)R_1(k_F,r)$ product
has its maximum at $r=0.7$ with value $0.393$,
and it becomes zero at $r$=2.53. For higher $r$ values it oscillates with
decreasing amplitudes around zero due to (shifted) Friedel oscillations. As Figure 3 shows,
the perturbative $j_0(k_F r)j_1(k_F r)$
product is much more {\it extended}. Its maximum (0.26) is at about $r$=1.7, it becomes zero
at about $r$=3.9, and has Friedel oscillations beyond this.
Clearly, the attractive proton is a strong perturbation at metallic densities.

\begin{figure}[tbp]
\resizebox{8.5cm}{!}{\includegraphics*{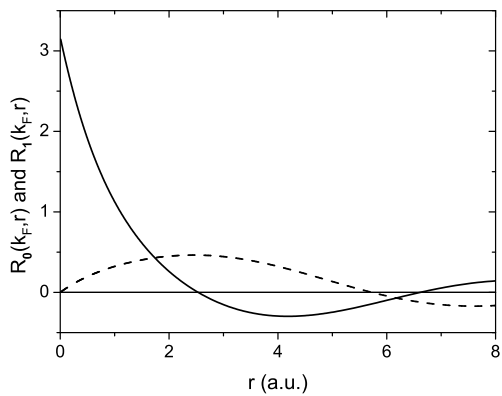}}
\caption{Self-consistently determined radial wave functions,
$R_0(k_F,r)$ and $R_1(k_F,r)$,
as a function of the radial distance $r$ in atomic units (a.u.). Solid and dashed
curves refer, respectively, to the $l=0$ and $l=1$ components.
The Wigner-Seitz parameter of the host system is $r_s=2.5$ (a.u.).}
\label{figure2}
\end{figure}
\begin{figure}[tbp]
\resizebox{8.5cm}{!}{\includegraphics*{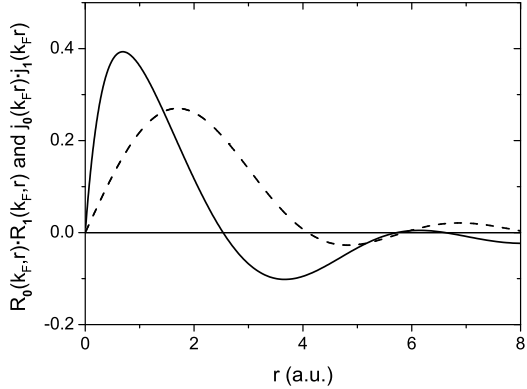}}
\caption{Products of the leading partial waves as a function of $r$. The solid and
dashed curves correspond, respectively,
to the self-consistent $R_0(k_F,r)R_1(k_F,r)$ and the perturbative $j_0(k_Fr)j_1(k_Fr)$.
The Wigner-Seitz parameter of the host system is $r_s=2.5$ (a.u.).}
\label{figure3}
\end{figure}

As a digression we enumerate at this point few important facts which support the reliability
of the above phase shifts and the radial $R_{l}(k_F,r)$ functions even in a real lattice
where the structural (atomic) surrounding could have influence. First, even unit charges ($Z=\pm{1}$)
represent strong {\it local} perturbations.
This is verified by experiments for slowing down of
{\it low}-speed ($v\, <\, v_F$) protons and antiprotons moving through paramagnetic
metallic targets characterized by ${3\pi^2} n_0=(p_F/{\hbar})^3$, where $p_F=m v_F={\hbar}k_F$.
In the theoretical \cite{Ferrell77} energy loss per unit path length
[$(dE/dx)=m v v_F n_0 {\sigma}_{tr}(p_F)$]
appears the transport-cross section

\begin{equation}
{\sigma}_{tr}(p_F)\, =\, \frac{4\pi}{(p_F/{\hbar})^2}\, \sum_{l=0}^{\infty}\, (l+1)\,
\sin^{2} [{\delta}_l(p_F) - {\delta}_{l+1}(p_F)].
\end{equation}
The theoretical results, based on $DFT$ phase shifts for protons
\cite{Echenique86} and antiprotons \cite{Nagy89}, are in impressive
agreement with $(dE/dx)$-data obtained \cite{Moller02} at dedicated
facilities of CERN and show, via the
${\sigma}_{tr}(Z=1,p_F)/{\sigma}_{tr}(Z=-1,p_F)$ ratio ($R$), a
pronounced ($R\geq{2}$) charge-sign effect which rules out
\cite{Nagy04} the applicability of a simple Born-approximation. For
completeness, for an antiproton one has ${\delta}_0(k_F)$=$-$0.7729
and ${\delta}_1(k_F)$=$-$0.2003 at $r_s=2.5$. The illustrative
Figures for the radial wave functions, and the above-outlined phase
shifts values for embedded unit charges ($Z=\pm{1}$) suggest [see
the discussion at Eq.(24) also] the use of the so-called
$sp$-approximation in order to get a reasonable estimation for our
coupling constant. This will be, therefore, the practical
approximation in the $TLS$ problem.

Quite remarkably, a {\it dimensionless} parameter
$K_{overlap}(d)$, which expresses \cite{Schonhammer91,Vladar93}
the overlap of two many-body ground states with the same
local potential at two sites separated by a short
distance $d$, depends on the phase shifts similarly

\begin{equation}
K_{overlap}(d)\, =\, \frac{1}{3}\, \left(\frac{2d}{{\lambda}_B}\right)^2\, \sum_{l=0}^{\infty}\, (l+1)\,
\sin^{2} [{\delta}_l(p_F) - {\delta}_{l+1}(p_F)],
\end{equation}
where ${\lambda}_B=2\pi{\hbar}/p_F$ is the de Broglie wavelength of an electron moving
with $v_F$ velocity; $(2d/{\lambda}_B)=(dk_F/{\pi})$.
Related to positive muon (${\mu}^{+}$) quantum diffusion in metallic targets, realistic phase shifts
were already applied \cite{Nagy99} to estimate this dimensionless parameter.

Second, while the band-structure paradigm emphasizes the importance of lattice structure,
calculation \cite{Jena79}
for the $Pd-H$ system using a molecular-cluster model shows that
the electronic properties of the impurity are dictated mainly by its {\it local} environment.
The such-calculated on-top screening density, ${\Delta}n(r=0)$, is only slightly smaller
than the simple jellium-based result which tends \cite{Almbladh76} to $1/{\pi}$ from above
as $n_0$ decreases.
Furthermore, the molecular-cluster and pseudo-jellium
calculations for $Pd-H$ are in close agreement with each other for $H$-displacements up to $0.5 a_0$ from
the equilibrium configuration. A self-consistent pseudopotential calculation
\cite{Louie83}
shows that the proton is screened on one atomic distance, and the hydrogen always has more
charge around it than the $Pd$ upto the Wigner-Seitz radius.
Note, that the amount of electron-localization in $H$-screening could characterize,
as was pointed out \cite{Vajeeston05} recently,
the site-preference of hydrogen in new storage materials.

We finish our supporting enumeration by associating the above {\it
local}-environment picture with the proposal of Hopfield
\cite{Hopfield69} for short-range properties when there is a change
in potential due to moving an atom by a small distance which results
in matrix-elements needed, in his case, to an estimation of
electron-phonon coupling in transition-metal superconductivity.
Namely, it was shown that {\it when} an angular momentum
decomposition of electron wave functions is used, the matrix
elements contain chiefly scatterings which change the angular
momentum of the electron. Precisely, it is this character which is
central in the context of scattering of electrons from two-level
systems. The parity change of the angular momentum state (without
altering the spin indices) of conduction electrons gives them an
internal degree of freedom coupled to that of the impurity
\cite{Zawadowski80,Emery92}. This internal degree is the background
to establish an analogy with the usual (spin-related, magnetic)
Kondo effect.

\newpage

$(ii)$ {\it The change of the potential for $d\ne{0}$.}
The screened charge is moved to one of the positions ${\sigma}^{z}=\pm{1}$ carrying the
potential, which is taken rigid as the screening action is very fast \cite{Borisov07}
compared to the infrared
processes essential in some other problems. The electron eigenfunctions are also carried with
the tunneling atom, but they are decomposed is terms of those have already been determined for the
central position; see in point $(i)$. Thus the new potentials are

\begin{equation}
V({\bf r}\, {\mp}\, {\sigma^z}\, \frac{d}{2}\, \hat z)\, =\, V({\bf r})\, {\pm}\,
{\sigma^z}\, {\Delta}V({\bf r}),
\end{equation}
where the important (perturbative at small $d$) term is given by

\begin{equation}
{\Delta}V({\bf r})\, =\, V({\bf r}\, - \frac{d}{2}\, \hat z)\, -\, V({\bf r}).
\end{equation}
A Taylor expansion in $d$ results in a simple dipolar form in the  leading order
\begin{equation}
\Delta\, V({\bf r})\, =\,  \frac{d}{2}\, \frac{{\partial}V(\bf r)}{{\partial}r_{\hat z}}\, =\,
\frac{d}{2}\, \cos{\theta}\, \frac{d V(r)}{d r}.
\end{equation}

We illustrate the behaviours of the self-consistent potential and its gradient in
Figure 4. Concretely, the $-rV(r)$ and $r^2[dV(r)/dr]$ products are plotted by solid
and dashed curves, respectively. The inset is devoted to the finer details of these
important functions. The calculated results refer to $r_s=2.5$ for the density parameter.
The remarkable Coulombic character of $r^2[dV(r)/dr]$ at {\it short} distance is due to a compensating
effect between the electrostatic (Hartree-term) screening and the local exchange-correlation term.

\begin{figure}[tbp]
\resizebox{8.5cm}{!}{\includegraphics*{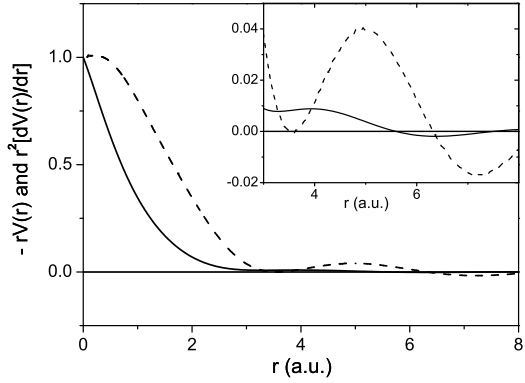}}
\caption{Characteristics of the self-consistent potential obtained at $r_s=2.5$ for the
screening of an embedded proton. The -$rV(r)$ and $r^2[dV(r)/dr]$ are plotted
as a function of $r$, by solid and dashed curves, respectively. 
The inset is devoted to finer details for $r>3$.
Atomic units are used.}
\label{figure4}
\end{figure}

In the following perturbation theories are applied in terms of ${\Delta}V({\bf r})$. 
The matrix elements are calculated between the states with wave functions
$\psi_{\bf k}^{\pm}({\bf r})$ determined in the previous section ($i$) but with the plane
wave functions are also presented. Namely, we use the distorted wave Born and the
usual Born approximations, respectively.

$(iii)$ {\it Matrix elements of the shifted potential between the
original wave functions.} 
In order to get a convenient, dimensionless \cite{Vladar83} coupling to 
characterize the effect of the perturbation on precalculated continuous 
states we shall use energy-normalization \cite{Friedrich06} for these:

\begin{equation}
\psi_{E}({\bf r})\, =\, \frac{1}{\sqrt{4\pi}}\, \sqrt{\frac{2mk}{\pi{\hbar}^2}}\, \psi_{{\bf k}}({\bf r}).
\end{equation}
The original matrix element at $k_F$ is multiplied, in such a way,
by the density of states ${\rho_0}(E_F)=(k_Fm)/(2{\pi}^2\hbar^2)=0.75
n_0/E_F$ per unit volume for a given spin evaluated at the Fermi
energy. Notice, oncemore, that the ${\Delta}F(k_1,k_2)$ {\it change}
in the scattering amplitude [i.e., the matrix element of the
perturbing ${\Delta}V({\bf r})$ between $\psi_{\bf k}^{\pm}({\bf
r})$ initial and final states \cite{Ballentine98}] should involve a
$m/(2\pi{\hbar}^2)$ prefactor.
With Eq.(13) the angle-integration over $\Omega_{\hat r}$ gives, by
applying standard recurrence relation for $\cos{\theta}\,
Y_{lm}(\theta,\phi)$, the following simple result

\begin{equation}
I(l,m)\, =\, \int\, d{\Omega_{\hat r}}\, \cos{\theta}\, Y_{lm}(\hat r)\, Y_{l'm'}^{*}(\hat r)\, =\,
\left[\frac{(l+1)^2 - m^2}{4(l+1)^2 - 1}\right]^{1/2}.
\end{equation}
Only the $l'=l+1$ and $m'=m$ values are allowed, due to the dipolar character.

{\it Born approximation}. It is instructive first to study the case
of a weak, $V_{ps}(r)$ pseudo-potential in Eq.(13). Thus, we perform
the radial integration in first-order Born approximation, i.e., we
apply plane-wave components for the free radial wave functions as

\begin{equation}
I_k(l)\, =\, \frac{2m}{\hbar^2}\, \int_{0}^{\infty}\, dr\, r^2\, \frac{dV_{ps}(r)}{dr}\, j_l({k}r)\, j_{l+1}({k}r).
\end{equation}
In this perturbative case the integration by parts and use of the following
expression based on recurrence relations for Bessel functions

\begin{equation}
\frac{d}{dr}\left[r^2j_l({k_1}r)j_{l+1}({k_2}r)\right]\equiv{r^2[k_2 j_l^2({k_1}r) - k_1 j_{l+1}^2({k_2}r)]},
\end{equation}
results in a remaining integral now with $V_{ps}(r)$.
In such a way we can apply the standard definition \cite{Ballentine98}
of the first-order Born ($B$) phase-shift

\begin{equation}
{\delta}_l^{B}(k)\, =\, - \frac{2mk}{\hbar^2}\, \int_0^{\infty}\, dr\, r^2\, V_{ps}(r)\, j_l^2({k}r),
\end{equation}
to obtain the following informative [see also Eq.(26), below]
expression

\begin{equation}
I_k^{B}(l)\, =\, \left[{\delta}_l^{B}(k)-{\delta}_{l+1}^{B}(k)\right].
\end{equation}
We stress that this equation is valid, physically, only for small values
of the phase shifts, i.e.,
when the distortion of the electron wave function by the central potential
field is negligable.

If a  weak potential is given via its Fourier representation, as in a dielectric screening
of the external field $V_{ps}(q)=V_{ext}(q)/{\epsilon}(q)$,
the Born phase shifts (we use atomic units here) are

\begin{equation}
{\delta}_l^{B}(k)\, =\, -\, \frac{1}{4{\pi}k}\, \int_0^{2k}\, dq\, q\, V_{ps}(q)\,
F\left(-l,l+1;1;\frac{q^2}{4k^2}\right),
\end{equation}
where $F$ is the standard hypergeometric function. In our simple,
so-called $sp$-approximation, one can get from Eq.(20) easily 
[see Eq.(24), also] the leading difference-term

\begin{equation}
{\delta}_0^{B}(k)-{\delta}_{1}^{B}(k)\, =\, \frac{1}{\pi}\, \frac{1}{(2k)^3}\,
\int_0^{2k}\, dq\, q^2\, [-q\,V_{ps}(q)].
\end{equation}
This $q$-space representation could be useful if the inverse
Fourier-Hankel transfomation results in a complicated function for
$V_{ps}(r)$, and a fast estimation is needed. For a commonly applied
Yukawa-type potential, $V^{Y}_{ps}(r)=-(Z/r)exp(-{\lambda}r)$, the
Born phase shifts from Eq.(18) are given by Legendre-functions
[$Q_l(x)$] of the second kind as

\begin{equation}
{\delta}_l^{BY}(k,\lambda)\, =\, \frac{Z}{k}\, Q_l(1+({\lambda}/k)^2/2).
\end{equation}
It is important to note, in the present context, that a restricted Friedel sum

\begin{equation}
Z\, =\, \frac{2}{\pi}\, \sum_{l=0}^{\infty}\, (2l+1)\, {\delta}_l^{BY}(k_F,\lambda)\,
\equiv{-\, \frac{k_F}{\pi^2}\, V^{Y}_{ps}(q=0)},
\end{equation}
still holds despite the perturbative approximation {\it if}
${\lambda}^{2}=4k_F/{\pi}$, i.e., the screening parameter
corresponds to the quasiclassical Thomas-Fermi value. Physically,
the normalization of the screening charge (calculated from the model
Yukawa form via Poisson's equation) is satisfied, but its real-space
distribution is not necessarily realistic.

The first-order Born phase shift does not contain the multiple
scattering effect in the central potential field. Formal
applications of Eqs.(18)-(19) with strong potentials might result in
an uncontrollable estimation for the effect in question. However,
the proposed correlation \cite{Zawadowski80} between the motion of
the TLS-atom and an angular-momentum change is now transparent even
at the first-order Born level, as Eq.(19) clearly shows. Note that
this transparency of correlation is obtained (easily) when one
applies \cite{Hopfield69} spherical-harmonics-based expansions for
the initial and final (plane waves in first-order Born
approximation) states separately, i.e., by implementing Eq.(4).

When, still in first-order Born approximation with a $V_{ps}(r)$ in
Eq.(13), we implement Eqs.(2-3) with the standard
spherical-harmonics expansion for $e^{-i{\bf q}{\cdot}{\bf r}}$,
i.e., without the mentioned {\it separation} for initial and final
states, one gets ($V^z_{ps}\equiv{V^z_{pw}}$) for small enough $d$

\begin{equation}
V^{z}_{pw}({\bf q})\, \sim{\, \frac{d}{2}\,
\int_0^{\infty}\, 4\pi\, r^2\,
j_1(qr)\, \frac{dV_{ps}(r)}{dr}\, dr}.
\end{equation}
In {\it this} representation of the (perturbative) matrix-element a
parity-change is not transparent. But a partial integration and the
use of the $j'_1(x)=j_0(x)-(2/x)j_1(x)$ recurrence relation gives
$-qV_{ps}(q)$ for the integral. The Fermi-surface average, ($q^2q\,
dq)\Rightarrow{[P_0(x)-P_1(x)]dx}$ with $x=\cos{\vartheta}$,
becomes, as Eq.(21) shows, a function of
[${\delta}^{B}_0(k_F)-{\delta}^{B}_1(k_F)$] {\it solely}. This
conclusion agrees with Ref.[6]; only the $l=0$ and $l=1$ harmonics
are relevant.

{\it Beyond the first-order Born approximation}. In a quite recent
theoretical work \cite{Zarand05}, which also rests on matrix element
calculation with a potential-gradient between normalized $s$ and $p$
{\it bound}-states, the bare [$V_{C}(r)=-Z/r$] Coulomb potential was
applied to characterize $TLS$ in a metallic matrix. Here, by using
the ${\eta}=Ze^2m/(k\hbar^2)$ Sommerfeld parameter, we add the
corresponding {\it exact} phase-shift difference for the case of a
Coulomb field

\begin{equation}
{\delta}_l^{C}(k)\, -\,{\delta}_{l+1}^{C}(k)\, =\, {\arctan}\left[\frac{\eta}{(l+1)}\right].
\end{equation}
By using the accurate $R_l({k},r)$ and $R_{l+1}({k},r)$ radial wave
functions to integration in Eq.(16) above [and not $R_l^2({k},r)$
and $R_{l+1}^{2}(k,r)$ to Eq.(18)] with gradient of the {\it true}
$V(r)$ behind {\it these} functions one obtains, based on earlier
result \cite{Gaspari72,Tang98}, the {\it exact}

\begin{equation}
I_k(l)\, =\, \sin[{\delta}_l(k)-{\delta}_{l+1}(k)]\, \equiv{
[\tan{\delta_l(k)}-\tan{\delta_{l+1}(k)}]\cos{\delta_l(k)}\, \cos{\delta_{l+1}(k)}},
\end{equation}
closed expression in terms of scattering phase shifts [see: Eqs.(9)
and (10), also]. The exact result, which contains now the multiple
scattering effect in the central potential field to all order, is a
bounded function in contrast to Eq.(19). We illustrate, in Figure 5,
the argumentum-function $r^2R_0(k_F,r)R_1(k_F,r)[dV(r)/dr]$ of
Eq.(16) by using the self-consistent solutions at $r_s=2.5$.
Fortunately, as noted above, we have already a simple, {\it
analytic} result in Eq.(26) for the integral without further
numerics.

\begin{figure}[tbp]
\resizebox{8.5cm}{!}{\includegraphics*{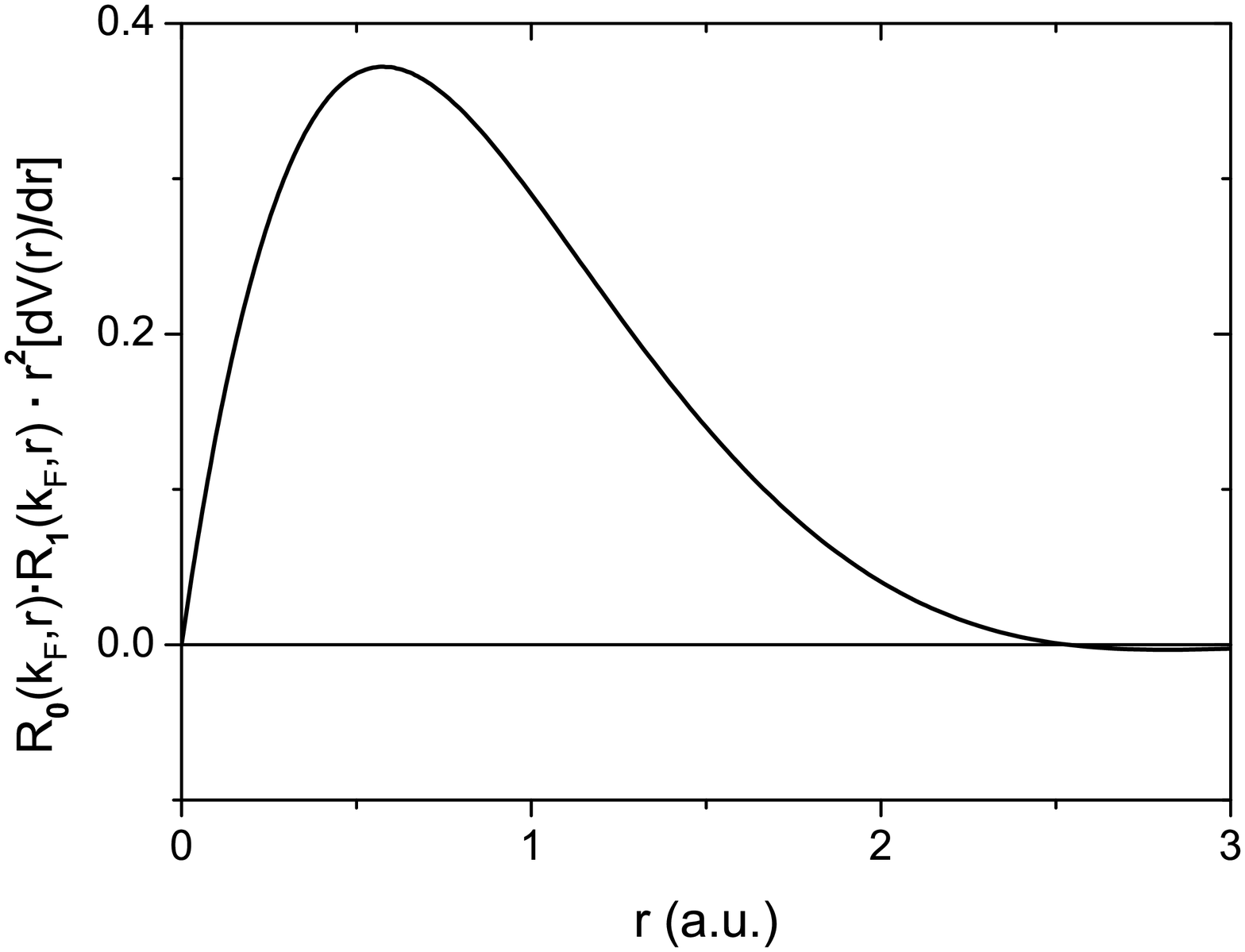}}
\caption{Illustration of the integrand of Eq.(16) obtained by using
self-consistent result for the potential gradient and the two
leading radial wave functions. The Wigner-Seitz parameter of the
host system is $r_s=2.5$. Atomic units are used.} \label{figure5}
\end{figure}
%


We add at this important point that with Eq.(25) and Eq.(26) one can
get the

\begin{equation}
I^{C}_k(l)\, =\, \frac{\eta}{[(l+1)^2 + {\eta}^2]^{1/2}},
\end{equation}
expression in terms of the parameters $Z$, $k$, and $l$, for a
Coulomb field; $I^{C}_{k}(l)\leq{1}$. The plane-wave-based
perturbative Coulomb limit, obtained with $V(r)=-Ze^2/r$ to Eq.(16)
is surprisingly similar to Eq.(27). The difference is only, but this
is crucial since $k_F\sim{1}$ in a.u., that there is {\it not}
$\eta^2$ term [see Eq.(21) also] in the denominator. Formal use of a
finite $l_{max}$ to Eq.(27) could mimic a screening-regularization
and, together with Eq.(15), could thus allow fast analytical
estimations. For the Coulomb case in the ${\eta}\rightarrow{\infty}$
limit (formally ${\hbar}\rightarrow{0}$) one has $[\delta^{C}_0(k) -
\delta^{C}_1(k)]\rightarrow {[{\pi}/{2} - {\hbar}v_e/{Ze^2}]}$,
where $v_e={\hbar}k/m$ is the electron velocity. Notice,
parenthetically, that this difference is one of the pedagogical
examples in physics which shows transparently the role of Planck's
constant, $\hbar$, "in action".

After the above detailed analysis on scattering, we return now to
the matrix element needed to a {\it suitable} \cite{Zawadowski80}
coupling introduced in Eq.(4). First we give, by using the previous
notations, a {\it dimensionless} expression related to the
scattering-amplitude change

\begin{equation}
{\Delta}f^z_{\alpha\beta}(k_F)
\equiv{(k_F/\pi)\, {\Delta}F_{{\alpha}{\beta}}^{z}(k_F)}\, = \,
\frac{d}{\lambda_B}\, I(l,m)\, I_{k_F}(l)\, A_{l+1}(k_F)\,
A_{l}(k_F),
\end{equation}
in which ${\alpha}$ and ${\beta}$ refer to the ${(l,m)}$ and
${(l+1,m)}$ values, respectively. Next, following earlier works
\cite{Zawadowski80, Vladar83}, we write a dimensionless form to the
{\it Hermitian} $H_1$ in Eq.(4) as

\begin{equation}
v_{{\alpha}{\beta}}^{z}(k_F)\, \equiv{\rho_0(E_F)\,
V^z_{\alpha\beta}(k_F)}.
\end{equation}
The obvious connection in the weak-perturbation limit is, based on
the peculiarity of the first-order Born ($B$) approximation with
plane waves, simply
$v_{{\alpha}{\beta}}^{z(B)}(k_F)={\Delta}f^{z(B)}_{\alpha\beta}(k_F)$.

This transparent connection between physical quantities at the Fermi
surface suggests us to use, beyond the above weak-coupling limit,
the $v_{{\alpha}{\beta}}^{z}(k_F)=Re\,
{\Delta}f^{z}_{\alpha\beta}(k_F)$ extension. With our choice for
boundary conditions to select initial and final states (involved in
matrix-element calculation based on the distorted wave Born method)
this seems to be the only logical step which preserves the important
Hermitian character of $H_1$ and reproduces the weak-coupling limit.
In addition, the proposed extension is in harmony with standard textbook 
statement \cite {Mahan00} on the characterization  of an energy {\it shift}
of a particle interacting with a potential.
Considering the above-deduced, remarkably simple rule on the true effect of an
angle-averaging at the Fermi surface [see, at Eq.(24)], we write

\begin{equation}
Re\, {\Delta}f^{z}_{00,10}(k_F)\, =\, \frac{1}{\sqrt3}\,
\frac{d}{{\lambda}_B}\, \sin[\delta_0(k_F)-\delta_1(k_F)]\,
\cos[\delta_0(k_F) + \delta_1(k_F)],
\end{equation}
\begin{equation}
Im\, {\Delta}f^{z}_{00,10}(k_F)\, =\, \frac{1}{\sqrt3}\,
\frac{d}{{\lambda}_B}\, \sin[\delta_0(k_F)-\delta_1(k_F)]\,
\sin[\delta_0(k_F) + \delta_1(k_F)]
\end{equation}
for the important real ($Re$) part, and the imaginary ($Im$) part
which is at least second-order in a weak-perturbation
($Z\rightarrow{0}$) limit.

Our physically-motivated extension to fix a value to the coupling of
a $TLS$ with conduction electrons, remains within the framework of
common knowledge: All stationary properties of metals which can be
described in terms of scattering of conduction electrons based on an
{\it adiabatic} picture, are periodic functions of the phase shift,
except, curiously, the Friedel sum rule. We stress that in our model
an electron merely sees a different scattering potential for each
state of the configuration but has no effective internal-spin degree
of freedom. The ${\psi}^{\pm}_{\bf k}$ states, for initial and final
states, are needed \cite{Ballentine98} to a perturbation theory in
the continuum of a given potential field. In experiments for
transport characteristics, we have current-carrying electrons. Thus
a standing-wave representation (which would refer to a different
boundary condition described by physics, for example by a cavity)
for the states involved in our matrix-element calculation is not
reasonable.

In the so-called unitary limit, ${\delta}_0(k_F)\sim{\pi/2}$, where
the effect of the self-consistent central $V(r)$ field is strong
[see, Figs. 2 and 3], the influence of the potential-shift
[${\Delta}V({\bf r})]$ becomes very small, i.e., $v^z_{00,10}(k_F)$
is small. Remarkably, this statement is in accord with the
conclusion of Gy\"orffy on a renormalized limit (obtained via
mapping to the partition function of a fictitious, auxiliary
logarithmic gas) with contact interaction \cite{Gyorffy78}. This
shows that the ground-state charge-distribution around an embedded
proton is, in fact, intact to small $(\sim{d})$ extra fields. In a
one-phase-shift approximation the maximal coupling would be at the
${\delta}_0(k_F)=\pi/4$ mathematical value. At this value the real
and imaginary parts of the scattering amplitude are equal in
magnitude. We shall return, briefly, to the simple contact-potential
approximation in the Appendix.

The expression in Eq.(28), which gives a strict linearity in $d$, is
based on a leading-term expansion for the perturbation
${\Delta}V(\bf r)$ as Eq.(13) shows. In order to get a more detailed
$d$-dependence of coupling due to the displacement of a screened
proton in the $z$-direction, we performed numerical
volume-integrations with Eq.(13) and the {\it dominating}
$R_0(k_F,r)Y_{00}(\hat r)R_1(k_F,r)Y_{10}(\hat r)$ product. By
introducing the simplifying notation

\begin{equation}
K^{z}(d)\, \equiv{|{\Delta}f_{{00},{10}}^{z}(k_F)|\, \lambda_B},
\end{equation}
motivated by the need (see below) of explicit estimation for a
physical observable, and performing the $\theta$-integration via a
variable-change, we have

\begin{equation}
K^{z}(d)\, =\,
\frac{2\sqrt3}{d^2} \int_0^{\infty} dr R_0(k_F,r)R_1(k_F,r) \int_{|r-d/2|}^{r+d/2}
du u V(u) [r^2 +(d/2)^2 -u^2].
\end{equation}

The numerical result for $K^{z}(d)$ is presented in Figure 6 by a solid curve. The dashed
curve refers to the asymptotic expansion, which is linear in $d$.
\begin{figure}[tbp]
\resizebox{8.5cm}{!}{\includegraphics*{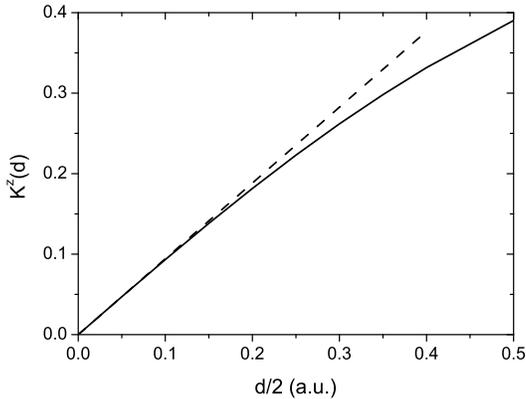}}
\caption{Numerical result (solid curve) for $K^{z}(d)$, defined in Eq.(33),
as a function of $(d/2)$. The dashed curve refers to the asymptotic Taylor expansion
which gives a linear dependence on $d$.
The Wigner-Seitz parameter of the host system is $r_s=2.5$. Atomic units are used.}
\label{figure6}
\end{figure}
We note that the asymptotic expansion provides a quite acceptable
representation upto about $d\simeq{0.5}$, and even at $d=1$ the deviation from the
numerical results is only about $25\%$.
The somewhat surprising linearity can be explained, partly, by the fact that
the second term ($\sim{\cos^2\theta}$) of a formal Taylor series
for the perturbation would give vanishing contribution, i.e., there is not quadratic,
$d^2$-proportional,
term at small $d$ in the $sp$ approximation.
Beyond the physically reasonable $d=1$ value, the $K^z(d)$ function grows more and more
gradually and it has a maximum at about $d\simeq{1.8}$. With an acceptable mathematical
accuracy one may use the $K^z(d)\simeq{0.52d\, (1+0.9d)\, exp({-0.9d})}$ fit for $d\leq{3}$
in our $sp$-dominated problem.

Finally, the magnitude of $|{\Delta}f^z_{00,10}(k_F)|$ can be
calculated by using our phase shifts at density parameter $r_s=2.5$
[in this case $k_F=(9\pi/4)^{1/3}/r_s\simeq{0.77}$], and assuming
(see Fig. 6) a size of the TLS $d=1$ and then
\begin{equation}
|{\Delta}f^z_{00,10}(k_F)|\, =\, \frac{k_F}{2\pi}\, K^z(d=1)\,
\simeq{0.05}.
\end{equation}
That value is moderate enough to ignore the multiple scattering
due to ${\Delta}V$ (see Appendix B), and
is in accordance \cite{Black79,Zawa83} with the {\it
estimated} typical values for different metallic glassy systems in
the intermediate coupling regions using ultrasound measurements. 
In our modelling of the amplitude-change due to a $TLS$, the absolute maximum at
$d\simeq{1.8}$ does not provide more than an about $25\%$ increase.
A naive {\it mixed} approximation, in
which the gradient of a bare Coulomb potential is weighted by our
self-consistent $R_0(k_F,r)R_1(k_F,r)$ product in Eq.(16), gives a
{\it smaller} numerical value than Eq.(26). This is due to,
mathematically, an over-weighting of the negative part of the
product in Fig. 3 by an unscreened-field gradient in the
volume-integral. A careful discussion of the underlying adiabatic
and anti-adiabatic pictures was given by Kagan \cite{Kagan86}, by
stressing the relevance of an adiabatic character due to the fast
screening-action, in $TLS$-motion.


\section{Summary and outlook}

A dimensionless  coupling constant that characterizes the effect of a potential-gradient perturbation
on scattering eigenstates of a self-consistently treated embedded impurity is deduced for small
values of the impurity
displacement $d$ in metallic electron gases. The result is expressed via bounded, trigonometric
functions of
scattering phase shift differences at the Fermi energy, quite similarly to a well-known overlap
parameter $K_{overlap}(d)$.
Beyond the leading-term expansion for the perturbation, the numerical results show that the
coupling parameter has an almost linear $d$-dependence upto the physically realistic $d=1 a_0$ value
for the displacement of a screened proton in an
electron gas with $r_s=2.5$.

As we mentioned earlier, at the enumeration of supporting facts
related to the applied self-consistent-field method, it is the
parity change of the angular momentum state of conduction electrons
which gives them an internal degree of freedom in the $TLS$ problem.
We derived this parity change in the present paper by using a
distorted wave Born approximation, i.e., using those scattering wave
functions to matrix-element calculation which are determined by the
central field of a fixed, spherically screened proton. Beyond the
applied distorted wave Born approximation, with an axially symmetric
scattering potential, the exact description would lead to {\it
coupled} radial equations \cite{Boardman67} in a partial wave
expansion of the scattered wave at a given $k$. Particularly, the
$d$-dependence of the coupling beyond the present renormalized
perturbation theory with precalculated spherical states, is an
exciting problem. Application of internally consistent (axially
symmetric) nonperturbative wave functions to the $TLS$ problem needs
a future study. The numerics could be based on a $DFT$-method
developed recently \cite{Muino00} for axially symmetric potentials
in an electron gas.

It is generally true that a clean theory, like the present one based
on scattering aspects, is an analysis of the properties of an
idealized, hypothetical model. What we have tried to do is to give a
self-contained mathematical treatment of a physically motivated
model, hydrogen in an electron gas, that can demonstrate the strong
local effect in an atomic displacement in a metallic target. A more
detailed, i.e., material-specific description of the host could be
based upon more realistic but much more complex physical
specifications. For example, simple application of a specified, as a
function of the hydrogen concentration, density of states at the
corresponding Fermi energy of a real $Pd-H$ system could change the
numerical value of the dimensionless coupling, found here with a
free-electron form  ${\rho_0}(E_F)=0.75 n_0/E_F$.

Our self-consistent Kohn-Sham approximation incorporates the
electron-electron interaction in the Hartree-like mean-field
potential which acts for independent electrons. Beyond this
approximation, and especially with bound states around the bare
impurity, the treatment of electron-electron interaction
(statistical and dynamical) has to be somewhat more sophisticated,
since part of it is already included in the screening. One must
avoid double counting, and consider self-interaction corrections. A
strong interaction between the electrons, involved in the screening
action, may well lead to the occurrence of localized magnetic
moments associated with an embedded impurity atom. In such a cases
the interplay between the orbital and spin degrees of freedom may
become an important question.

As we fixed in the present paper, in our model an electron merely
sees a different scattering potential for each state of the
(impurity) configuration but has no effective internal-spin degree
of freedom. A nontrivial extension of the applied method, i.e., the
distorted wave Born approximation, could be a problem where in
addition to our displacement ${\Delta}V({\bf r})$ there is an extra
spin-orbit ($so$) coupling already at the level of our $H_0$ via a
scattering interaction: $V(r)\Rightarrow{V(r)+V_{so}(r)}$, where
$V_{so}(r)\sim{[r^{-1}V^{'}(r)\, \bm {L}\cdot\bm{\sigma}]}$. As was
demonstrated \cite{Ballentine98} by Ballentine, the {\it change} in
the scattering amplitude
has a peculiar character in this case. Namely, there is no term with
$l=0$ in the amplitude-change, because the operator ${\bf L}$ yields
zero in that case. How a combined ${\Delta}V({\bf r})+V_{so}(r)$
perturbation could change the statement obtained at Eq.(24) in a
$TLS$ problem might deserve a detailed study.

The theoretical modeling of important observables based on
transport-related experiments needs additional care. For example, a
recent calculation shows the possibly important role of different
charge-states in current-driven electromigration and backflow,
\cite{Vincent07} where the long-range characters of scattered waves
are more important than in our present problem for a $TLS$ coupling.
We have discussed and emphasized the importance of physical boundary
conditions (which appear in the fundamental Lippmann-Schwinger
equation) for continuous states which are based on a second-order
differential (Schr\"odinger) equation. If our displacing-atom were
embedded into a system described by externally generated
\cite{Limot05} standing waves, one could use \cite{Mahan00} the
$e^{i{\delta}_l(k)}\Rightarrow{1/\cos{\delta}_l(k)}$
normalization-change and, thus, get for the right-hand-side of
Eq.(26) {\it only} the difference of $tangents$. In this case, the
coupling could enhance almost resonantly. Indeed, the mentioned
$STS$ study \cite{Limot05} heralds, via spectroscopic informations,
a common nonmagnetic effect with different impurities in
standing-wave patterns. We speculate that the dissolution of
hydrogen into $Pd$-electrodes can make a local confinement-like
effect for states involved in the conductance \cite{Csonka04} of
$Pd-H$ nanojunctions.

Finally, a way to consider electronic inhomogeneities can be an {\it
additional} local-density approximation governed by the strong
short-range distortion on which our model is based. Between $Pd$
atoms of a real lattice there are ranges, measured from a
lattice-atom position, where the density of states has enhanced
local values. If the allowed tunneling occurs in such geometrical
ranges, a mathematical averaging of our $v^z[r_s(r)]$ function over
a certain range of $r_s(r)$ seems to be reasonable. The role of an
almost ferromagnetic nature of a real $Pd$ target (expressed via a
Stoner-enhancement \cite{Suhl75} in the spin response function)
might also deserve future considerations. These combined
considerations could give further quantitative information to a
field of considerable experimental relevance.


{\bf Acknowledgments}. 
The authors thank L. Borda, R. Diez Mui\~no,
and G. Zar\' and for useful discussions. A.Z. is grateful to the
Humboldt Foundation to support his stay in Munich, where part of
this work was done. I.N. acknowledges the warm hospitality at the
DIPC, San Sebasti\'an. The work has been supported partly by the
OTKA: Grant Nos. T046868 and T049571 for I.N., and Grant Nos.
T048782 and TS049881 for A.Z., respectively.


\section{Appendix A: Connection with earlier theories}

This Section is devoted to a comparison with previous works
\cite{Vladar83,Zarand95,Cox98} on modeling the coupling in Eq.(2),
based on plane-wave states. In order to provide a clear
phenomenology, we stress the point that the present theoretical
description, which is based on calculation of matrix-elements of a
dipolar potential-perturbation, also uses prefixed basis sets.
Namely, plane-waves in the conventional and precalculated
self-consistent ones in the distorted-wave Born approximation for
continuous states.

In the main text, at the details of the first-order Born
approximation, we outlined the $q$-representation of the coupling.
Thus the desired link, for example to Ref.[46] with a Yukawa
potential, is easily obtained by using the scattering value of
$q=2k_F\, {\sin}({\vartheta}/2)$ and the

\begin{equation}
V_{ps}(q)\, =\, \frac{4{\pi}Ze^2}{q^2 + {\lambda}^2}\equiv{\frac{2\pi\hbar^2}{mk_F}\, \sum_{l=0}^{\infty}
(2l+1){\delta}^{B}_{l}(k_F)\, P_l(\cos{\vartheta})},
\end{equation}
Born-representation to Eq.(24) for the $V^z({\bf q})$ quantity;
$P_l(\cos{\vartheta})\sim{\sum_{m}\, Y^{*}_{lm}({\hat
k_1})Y_{lm}({\hat k_2})}$. Motivated by certain scattering-length
arguments, the $a_l$ notations were adopted in the mentioned earlier
works instead of the ${\delta}^{B}_l(k)/k$ ratios. Somewhat
fortuitously, the numerical value of the perturbative difference in
Eq.(19) with a Yukawa-type (${\lambda}^2=4k_F/{\pi}$) screening

\begin{equation}
{\delta}^{B}_0(k_F) - {\delta}^{B}_1(k_F)\, =\,
\frac{Z}{k_F}\left[1 - \left(\frac{\lambda}{2k_F}\right)^2\,
ln\left(1 + \frac{4k_F^2}{\lambda^2}\right)\right],
\end{equation}
is not far, at least for $Z=1$, from the precise value obtained from
Eq.(26) with self-consistently determined phase shifts of an
embedding problem. Qualitatively, the mistake one makes in choosing
a linearly screened potential is "compensated" by the use of the
first-order Born approximation; neither of these approximations are
quantitatively accurate.

More importantly, the careful numerical analysis of Ref.[46],
performed by {\it assuming} higher ($l>1$) angular-momentum channels
(i.e., using more $a_l$ parameters) to an estimation on
$TLS$-coupling, shows that (contrary to naive earlier expectations)
there {\it is not} a series of Kondo-like effects corresponding to
the increase of different orbital channels. This, numerics-based
statement, is in harmony with the rule established at Eq.[24] in the
present paper.

A second-order Born approximation [valid for $(Z/k_F)<1$]
for the scattering (transition) amplitude [$f^{(2)}$] with the above
simple Yukawa potential gives

\begin{equation}
2{\pi} f^{(2)}(q,k_F)\, =\, \, V^{Y}_{ps}(q)\, +\, \frac{4{\pi}Z^2{\lambda}}
{{\lambda}^4 + 4k_F^2{\lambda}^2 + k_F^2q^2},
\end{equation}
in atomic units, for simplicity. This approximation with a
fixed, linear-response-based input potential would suggest an enhancement of the coupling.
A more consistent ($cons$) treatment, in which the linear screening is also modified upto
the second-order by a quadratic-response method, reduces \cite{Nagy04}
this enhancement as follows

\begin{equation}
f_{cons}^{(2)}(q,k_F)\, =\, f^{(2)}(q,k_F)\, -\,
\frac{2 Z^2{\lambda}}
{{\lambda}^4 + 4k_F^2{\lambda}^2 + (4k_F^2 + {\lambda}^2)q^2},
\end{equation}
showing that $f^{(2)}_{cons}(q,k_F)>V^{Y}_{ps}(q)/{2\pi}$ still holds for $q\neq{0}$, but
in the forward (${\vartheta}=0$) limit $f^{(2)}_{cons}(q=0,k_F)=V^{Y}_{ps}(q=0)/{2\pi}$.
This observation, which is based on selected (RPA) diagrams, heralds that care is
needed when one uses a higher-order method
in terms of a bare (in our case: a linearly screened) input potential in
field-theoretic many-body attempts.

A central local potential, for example $V_{ps}(r)$, gives rise to
scattering of all orders of spherical harmonics even at the
first-order Born level. On the other hand, a so-called separable
potential \cite{Yamaguchi54,Nozieres69} for a given orbital ($l$)
momentum causes scattering only for the given ($l$th) partial wave;
in the case of all $l$, one speaks of a completely separable
potential. A correct determination of the corresponding
channel-potentials,
${\bar{V_l}}(k_1,k_2)={\bar{V}}_lu_l(k_1)u_l(k_2)$, in this modeling
could rest on experimental data or on a detailed, microscopic theory.

It is important from the point of view of physical consistency, that
the frequently applied tangent-method \cite{Kohn51}
on multiple scattering effects gives \cite{Nozieres69} a
$\tan{\bar{\delta}}_0(k_F)$ in terms of $k_F$, $u_0(k_F)$, and
${\bar{V}}_0$ of the $s$-channel potential. In the even simpler case
with a contact $V_0({\bf r})\Rightarrow{{\bar{V}}_0{\delta({\bf r})}}$ auxiliary interaction 
(which gives a constant potential in momentum space) 
one has the remarkably simple form
$\tan{\bar{\delta_0}}=-\pi{\rho}_0\bar{V}_0$ for the $s$-wave phase shift. 
Note that by writing, formally, $\tan{\bar{\delta}_0}$ in the
lhs of Eq.(18), one can get this exact result in one-step from

\begin{equation}
\tan{\bar{\delta}_0}\, =\, -\, \frac{2mk_F}{{\hbar}^2}\, \int_{0}^{\infty}\,
dr\, r^2\, \left[\frac{\bar{V}_0}{4{\pi}r^2}\, {\delta}(r)\right]\, j_{0}^2(k_Fr).
\end{equation}
The {\it precise} derivations of the exact result rest on much more involved
calculations by using real-space \cite{Calarco06} or momentum-space \cite{Yu84} 
Schr\"odinger equation with ${\bar{V}}_0{\delta}({\bf r})$.
Application of a renormalized contact interaction, defined via
$-\bar{V}_0\Rightarrow{[1/(\pi{\rho}_0)]\tan{\bar{\delta}_0}}$, in the perturbative
rhs of Eq.(39) results in an identity. Clearly, this renormalized interaction could
be used with unperturbed plane wave states to calculate the real reactance-matrix.
The exact standing ($st$) wave solution [cf. Eq.(45)] of the tangent-method is the following

\begin{equation}
{\phi}_{st}(k_F,{\bf r})\, =\, \frac{1}{\cos{\bar{\delta}_0}}\,
\frac{\sin(k_Fr + \bar{\delta}_0)}{k_Fr}.
\end{equation}
This is based, oncemore, on the principal-value Green's function.

The standard logic to determine a value of $\bar{V}_0$  
is based on the phase shift [${\delta}_0(k)$] of the real potential [$V_0(r)$],
but the $s$-channel contact interaction \cite{Borda03} has an important
limitation when we apply it to the screening problem of a charge
$Z$. It was pointed out earlier \cite{Clogston62} that this model
cannot supply enough charge to shield the Coulomb field of the
physically simplest impurity $Z=1$. A {\it formal} requirement of
${\delta}_0(k_F)={\bar{\delta}_0}$ would result in
$\bar{V}_0\rightarrow{\infty}$, when the self-consistent
[${\delta}_0(k_F)$] leading phase shift (for $Z=1$) goes to
${\pi}/2$. At $r_s=2.5$ of the Wigner-Seitz density parameter, an
about $\bar{V}_0\simeq{20}$ value (in a.u.) is prescribed by this
formal constraint. 

The auxiliary contact-potential was applied earlier within a local-phonon model of 
electron-phonon interactions \cite{Yu84} in an opposite way, i.e., via a direct
approximation ($\bar{V_0}\sim{d}$) for it to characterize the (constant) 
momentum-space potential of an atomic displacement
in a free electron gas. In this case $\tan{\bar{\delta}_0}$ 
could measure a renormalized effect of a prefixed input $\bar{V}_0$,
beyond the conventional first-order Born approximation 
($\bar{\delta}^{(B)}_0=-\pi{\rho}_0\bar{V}_0$). 
Motivated by that work, the scaling equations of the two-level problem has 
been investigated \cite{Vladar86, Vladar88} 
in details in terms of an input $\bar{\delta}_0\leq{\pi/2}$ variable
in the so-called large-phase-shift case.
The mentioned direct approximation was implemented \cite{Zarand05} recently as
$\tan{\bar{\delta_0}}\equiv{\pi\, v^{z}}$ with a dimensionless matrix
element ($v^z\sim{d}$) of the Coulomb-potential gradient of an embedded proton
taken between the corresponding hydrogenic ($1s$ and $2p$) bound states, and
weighted by an enhanced density of states due to an other 
model potential; cf. Ref.[55].

The scattering with contact interaction is isotrop, i.e., the scattering amplitude 
(and thus the diagonal transition $T$-matrix)
does not depend on the scattering angle.
The corresponding, dimensionless [$k_F \bar{F}_0(k_F)$] exact scattering amplitude is

\begin{equation}
k_F \bar{F}_0(k_F)\, =\, e^{i\bar{\delta_0}}\, \sin{\bar{\delta_0}}\, =\, 
-\, \frac{\tan{\bar{\delta}_0}}{1 - i\tan{\bar{\delta}_0}}.
\end{equation}
Its real part has the simple form of

\begin{equation}
k_F\, Re \bar{F}_0(k_F)\, =\, \pi{\rho}_0\, \frac{\bar{V}_0}{1 + (\pi{\rho}_0\bar{V}_0)^2},
\end{equation}
which could suggest, based on a proper reinterpretation of the rhs, an effective (contact) potential
to calculations with unperturbed plane wave states. 
If the impurity scattering is purely local one can use
a many-body Green-function-based method \cite{Doniach74} also
to derive the exact result in Eq.(41), 
as the $T$-matrix depends only on energy and not on momentum; 
such a derivation shows the algebraic nature of the propagator method.

For a single impurity embedded in an electron gas, and described by a  
regular $V_0({\bf r})$ potential,
the electron Green's function can be written symbolically as

\begin{equation}
G\, =\, G_0 + G_0\, T\, G_0,
\end{equation}
using the standard $T$-matrix approach \cite{Doniach74}. In our case
the bare Green's function of a free electron of the ideal
system with chemical potential $\mu$
is $G_0(\omega,{\bf k})=[{\omega}-({\epsilon}_{\bf k}-\mu)+i0]^{-1}$.
Since the Fourier transform of a {\it contact} interaction is $\bar{V}_0$, 
the equation for the corresponding $T$-matrix is particularly simple as it depends only
on energy and not on momentum

\begin{equation}
T(\omega)\, =\, \bar{V}_0\, +\, \bar{V}_0\, \sum_{{\bf k}}G_0(\omega,{\bf k})\, T(\omega).
\end{equation}
One easily gets the $T={\bar{V}_0/(1-\bar{V}_0 \bar{G}_0)}$ solution, where
the $\bar{G}_0(\omega)$ quantity is the local propagator at the impurity site averaged
in the momentum-vector and taken at the Fermi surface, $|{\bf k}|=k_F$ and ${\omega}=0$.
Thus one has \cite{Doniach74} the simple $\bar{G}_0=-i{\pi}{\rho}_0$ relation,
which results in Eq.(41) for the complex scattering amplitude; $\pi{\rho}_0=mk_F/(2\pi\hbar^2)$.
Finally, the exact solution for the scattered wave, with $V_0({\bf r})=\bar{V}_0\delta({\bf r-r_0})$ and
at $k=k_F$, is given by

\begin{equation}
{\psi}_{k_F}^{+}({\bf r})\, =\, e^{i\bar{\delta}_0}\, \frac{\sin(k_F|{\bf r-r_0}| + \bar{\delta}_0)}{k_F|{\bf r-r_0}|},
\end{equation}
showing a limited applicability to a {\it realistic} enhancement at the impurity position.

The main point in such a standard calculation is to show that the Green function
method can, sometimes, demonstrate very general properties of many-body systems;
here only the density of states of the unperturbed system comes into the final answer in Eq.(44). 
It is, on the other hand, not necessarily the easiest way of calculating beyond the 
{\it contact-form} for an impurity potential. The propagator, being essentially a function of two
vector variables, is inevitably more complicated mathematically than a one-particle wave function.


\section{Appendix B: Multiple scattering off a localized ${\Delta}V$}

Until now the assumption has been used that the perturbation due to ${\Delta}V$ is weak, 
which is certainly the case for small $d$ and valid in the parameter range has been discussed.
The effect of this perturbation was considered within a distorted-wave Born approximation, i.e., via the
{\it first} term of the distorted-wave Born series; the outgoing state of $V$ is used for the
final state of $V+{\Delta}V$. A precise treatment for arbitrary $d$, as we mentioned in Sec. III,
should rest on a set of coupled radial equations of an axially symmetric potential.
Such a treatment could model faithfully the $d$-dependence of coupling due to atom-displacements.

For completeness, it is worthwile to discuss the multiple scattering in ${\Delta}V$ which is also
makes connection to some related earlier works \cite{Yu84,Vladar86,Vladar88}. 
We outline, therefore, a Green-function-based analysis, supposing a convenient {\it strongly 
localized} potential-form and convergency of the corresponding distorted-wave Born series. 
Using a schematical notation, $V + {\Delta}V$, the Dyson equation for the total 
Green's function $G$ is as follows

\begin{equation}
G\, =\,  G_0\, + G_0\, (V + {\Delta}V)\, G
\end{equation}
where $G_0$ refers to the free electron. That equation can be solved in two steps.
First, the summation is made only in $V$ and an intermediate Green's function $G^{V}$ 
(corresponding to propagation in $V$) is defined symbolically as 

\begin{equation}
G^{V}\, =\,  \frac{G_0}{1 - V G_0}.
\end{equation}

That can be expressed \cite{Ballentine98} by the exact independent-electron wave functions ${\psi}^{+}_k({\bf r})$,
given by Eq.(6) in the presence of the self-consistent $V$  in an electron gas as

\begin{equation}
G^V({\bf r},{\bf r'},\omega)\, =\, 
{\int}\frac{d{\bf k}}{(2\pi)^3}\,
\frac{[{\psi}_{\bf k}^{+}({\bf r'})]^{*}{\psi}_{\bf k}^{+}({\bf r})}{\omega -({\epsilon}_k-\mu) +i0}. 
\end{equation}
After performing the integration with respect to the direction of ${\bf k}$, one gets

\begin{equation}
G^V({\bf r},{\bf r'},\omega)\, =\, \frac{1}{2\pi^2}\, \sum_{lm}\, 
{\int}\, dk\, k^2\,
\frac{{\varphi}_{klm}^{*}({\bf r'}){\varphi}_{klm}({\bf r})}{\omega -({\epsilon}_k-\mu) +i0}, 
\end{equation}
where ${\varphi}_{klm}=R_l(k,r)Y_{lm}({\hat r})$.  
In the usual ($m=0$) approximation this equation is simplified (see Appendix A) and results in 

\begin{equation}
G^V({\bf r},{\bf r'},{\omega}=0)\, =\, - i\pi\rho_0\, \sum_{l}\varphi_{k_Fl}({\bf r'})
\varphi_{k_Fl}({\bf r}),
\end{equation}
which is a generalization of the noninteracting Green function based on plane waves.

Second, the term ${\Delta}V$ is taken into account, in analogy of Eq.(43), as

\begin{equation}
G\, =\, \frac{G^V}{1 -{\Delta}VG^V}\, =\, G^V + G^V{\Delta}V_{eff}G^V,
\end{equation}
where a nonlocal effective ({\it eff\,}) term is introduced as 

\begin{equation}
{\Delta}V_{eff}\, =\, \frac{{\Delta}V}{1 - {\Delta}VG^V}.
\end{equation}
In the general case Eq.(51) in real space is very complex, thus only the 
special case of a well-localized ${\Delta}V$ is treated;
$G^{V}(0,0,0)=-i\pi{\rho_0}E(k_F)$ where $E(k_F)$ is an enhancement due to $V$.
Thus, the matrix elements of a strongly localized ${\Delta}V$ are calculated. 

The wave functions ${\varphi}_{kl}({\bf r})$
are regular (see, Fig. 2), thus they can be expanded in ${\bf r}$ and only 
the first corrections are kept ($k_Fr\ll{1}$). With such constraints one gets

\begin{equation}
\varphi_{k_Fl}({\bf r})\, =\, \delta_{l,0}\, \varphi_{k_Fl}({\bf r}=0) +
\delta_{l,1}\, {\bf r}\, {\cdot}\, \frac{\partial}{\partial {\bf r}}\varphi_{k_Fl}({\bf r}=0).
\end{equation}
The matrix of ${\Delta}V$ can be calculated as e.g.

\begin{equation}
{\Delta}V_{1,0}\, =\, 
\varphi_{k_Fl=0}({\bf r}=0)\, {\int}
\left(\frac{\partial\varphi_{k_Fl=1}}{\partial {\bf r}}\, \right)_{{\bf r=0}}\, {\bf r}\,
{\Delta}V({\bf r})\, d{\bf r}.
\end{equation}
Note that ${\Delta}V_{1,0}={\Delta}V_{0,1}$. Furthermore, ${\Delta}V_{00}=0$
as the volume integral of a dipolar ${\Delta}V({\bf r})$ is zero,
and ${\Delta}V_{11}$ is $O(d^2k_F^2)$.
In such a way the matrix element of the effective potential can be 
[${\pi}{\rho}_0=k_F/(2\pi)]$ determined as

\begin{equation}
({\Delta}V_{eff})_{1,0}\,=\, \frac{{\Delta}V_{1,0}}{1 + {\pi}^2{\rho}_0^2({\Delta}V_{1,0})^2}.
\end{equation}
Therefore, for the physically realistic small $d$ (where ${{\Delta}V}\sim{d}$) the above-outlined 
treatment heralds a third-order change in $d$ beyond the $d$-linear term. 
Interestingly, the {\it form} in Eq.(55) is quite similar to Eq.(42).
Finally, as we discussed in Appendix A, a convenient phase shift ($\bar{\delta}_0$) could be
introduced as $\bar{\delta}_0=-\arctan(\pi{\rho}_0{\Delta}V_{0,1})$.



\begin{thebibliography}{00}
%
\bibitem{Beck81}
J. L. Black, {\it Metallic Glasses}, edited by H. J. G\" untherodt and H. Beck
(Springer Verlag, Berlin, 1981), p. 167.
%
\bibitem{Black79}
J. L. Black, B. L. Gy\" orffy, and J. J\" ackle, Philos. Mag. B {\bf 40}, 331 (1979).
%
\bibitem{Cochrane75}
R. W. Cochrane, R. Harris, R. O. Strom-Olsen, and M. J. Zuckerman, Phys. Rev. Lett. {\bf 35},
676 (1975).
%
\bibitem{Zawadowski80}
A. Zawadowski, Phys. Rev. Lett. {\bf 45}, 211 (1980).
%
\bibitem{Vladar83}
K. Vlad\' ar and A. Zawadowski, Phys. Rev. B {\bf 28}, 1564 (1983).
%
\bibitem{Aleiner01}
I. L. Aleiner, B. L. Altshuler, Y. M. Galperin, and T. A. Shutenko,
Phys. Rev. Lett. {\bf 86}, 2629 (2001).
%
\bibitem{Zarand05}
G. Zar\' and, Phys. Rev. B {\bf 72}, 245103 (2005).
%
\bibitem{Ralph92}
D. C. Ralph and R. A. Buhrman, Phys. Rev. Lett. {\bf 69}, 2118 (1992).
%
\bibitem{Cox98}
For a review, see: D. L. Cox and A. Zawadowski, Adv. Phys. {\bf 47}, 599 (1998).
%
\bibitem{Borda03}
L. Borda, A. Zawadowski, and G. Zar\' and, Phys. Rev. B {\bf 68}, 045114 (2003).
%
\bibitem{Csonka04}
Sz. Csonka, A. Halbritter, G. Mih\' aly, O. I. Shklyarevskii, S. Speller, and
H. van Kempen, Phys. Rev. Lett. {\bf 93}, 016802 (2004);
A. Halbritter, L. Borda, and A. Zawadowski, Adv. Phys. {\bf 53}, 939 (2004).
%
\bibitem{Djukic05}
D. Djukic, K. S. Thygesen, C. Untiedt, R. H. M. Smit, K. W.
Jacobsen, and J. M. van Ruitenbeek, Phys. Rev. B {\bf 71}, 161402(R)
(2005).
%
\bibitem{Ballentine98}
L. E. Ballentine, {\it Quantum Mechanics} (World Scientific, Singapore, 1998);
G. R. Satchler, {\it Direct Nuclear Reactions} (Clarendon Press, Oxford, 1983).
%
\bibitem{Peierls79}
R. Peierls, {\it Surprises in Theoretical Physics} (Princeton
University Press, Princeton, 1979), Sec. 6.3.
%
\bibitem{Morel62}
P. Morel and P. W. Anderson, Phys. Rev. {\bf 125}, 1263 (1962).
%
\bibitem{Ashcroft04}
N. W. Ashcroft, Phys. Rev. Letters {\bf 92}, 187002 (2004).
%
\bibitem{Kohn99}
W. Kohn, Rev. Mod. Phys. {\bf 71}, 1253 (1999).
%
\bibitem{Almbladh76}
C. O. Almbladh, U. von Barth, Z. D. Popovic, and M. J. Shore, Phys. Rev.
B {\bf 14}, 2250 (1976); E. Zaremba, L. M. Sander, H. B. Shore, and J. H. Rose,
J. Phys. F: Metal Phys. {\bf 7}, 1763 (1977).
%
\bibitem{Perrot94}
F. Perrot and M. Rasolt, J. Phys.: Condens. Matter {\bf 6}, 1473 (1994).
%
\bibitem{Ferrell77}
T. L. Ferrell and R. H. Ritchie, Phys. Rev. B {\bf 16}, 115 (1977).
%
\bibitem{Echenique86}
P.M. Echenique, R.M. Nieminen, J.C. Ashley, and R.H. Ritchie, Phys. Rev.
A {\bf 33}, 897 (1986).
%
\bibitem{Nagy89}
I. Nagy. A. Arnau, P. M. Echenique, and E. Zaremba, Phys. Rev. B {\bf 40}, R11983 (1989); {\it ibid}\,
{\bf 44}, 12172 (1991).
%
\bibitem{Moller02}
S. P. Moller, A. Csete, T. Ichioka, H. Knudsen, U. I. Uggerhoj, and H. H. Andersen,
Phys. Rev. Lett. {\bf 88}, 193201 (2002).
%
\bibitem{Nagy04}
For a review, see: I. Nagy and B. Apagyi, Adv. Quantum Chem. {\bf 46}, 268 (2004).
%
\bibitem{Schonhammer91}
K. Sch\" onhammer, Phys. Rev. B {\bf 43}, 11323 (1991).
%
\bibitem{Vladar93}
K. Vlad\' ar, Prog. Theor. Phys. {\bf 90}, 43 (1993).
%
\bibitem{Nagy99}
I. Nagy, B. Apagyi, J. I. Juaristi, and P. M. Echenique, Phys. Rev. B {\bf 60}, R12546 (1999).
%
\bibitem{Jena79}
P. Jena, F. Y. Fradkin, and D. E. Ellis, Phys. Rev. B {\bf 20}, 3543 (1979).
%
\bibitem{Louie83}
C. T. Chan and S. G. Louie, Phys. Rev. B {\bf 27}, 3325 (1983).
%
\bibitem{Vajeeston05}
P. Vajeeston, P. Ravindran, R. Vidya, A. Kjekshub, and H. Fjellvag,
Europhys. Lett. {\bf 72}, 569 (2005).
%
\bibitem{Hopfield69}
J. J. Hopfield, Phys. Rev. {\bf 186}, 443 (1969).
%
\bibitem{Emery92}
V. J. Emery and S. Kivelson, Phys. Rev. B {\bf 46},  10812 (1992).
%
\bibitem{Borisov07}
M. Quijada, A. G. Borisov, I. Nagy, R. Diez Mui\~no, and P.M. Echenique,
Phys. Rev. A {\bf 75}, 042902 (2007).
%
\bibitem{Friedrich06}
H. Friedrich, {\it Theoretical Atomic Physics} (Springer, Berlin, 2006), Sec. 1.3.4.
%
\bibitem{Gaspari72}
G. D. Gaspari and B. L. Gy\" orffy, Phys. Rev. Lett. {\bf 28}, 801 (1972).
%
\bibitem{Tang98}
J. M. Tang and D. J. Thouless, Phys. Rev. B {\bf 58}, 14179 (1998).
%
\bibitem{Mahan00}
G. D. Mahan, {\it Many-Particle Physics} (Plenum, New York, 2000).
%
\bibitem{Gyorffy78}
J. L. Black and B. L. Gy\"orffy, Phys. Rev. Letters {\bf 41}, 1595
(1978).
%
\bibitem{Zawa83}
K. Vlad\' ar and A. Zawadowski, Phys. Rev. B {\bf 28}, 1596 (1983).
%
\bibitem{Kagan86}
Yu. Kagan and N. V. Prokof'ev, Sov. Phys. JETP {\bf 63}, 1276 (1986).
%
\bibitem{Boardman67}
A. D. Boardman, A. D. Hill, and S. Sampanthar, Phys. Rev. {\bf 160}, 472 (1967).
%
\bibitem{Muino00}
R. Diez Mui\~no and A. Salin, Phys. Rev. B {\bf 60}, 2074 (1999); {\it ibid}\,
{\bf 62}, 5207 (2000).
%
\bibitem{Vincent07}
R. P. Vincent, I. Nagy, and E. Zaremba, Phys. Rev. B {\bf 76},
073301 (2007).
%
\bibitem{Limot05}
L. Limot, E. Pehlke, J. Kr\"oger, and R. Berndt, Phys. Rev. Letters
{\bf 94}, 036805 (2005).
%
\bibitem{Suhl75}
E. G. d'Agliano, P. Kumar, W. Schaich, and H. Suhl, Phys. Rev. B {\bf 11}, 2122 (1975).
%
\bibitem{Zarand95}
G. Zar\'and, Phys. Rev. B {\bf 51}, 273 (1995).
%
\bibitem{Yamaguchi54}
Y. Yamaguchi, Phys. Rev. {\bf 95}, 1628 (1954).
%
\bibitem{Nozieres69}
P. Nozieres and C. T. De Dominicis, Phys. Rev. {\bf 178}, 1097 (1969).
%
\bibitem{Kohn51}
W. Kohn, Phys. Rev. {\bf 84}, 495 (1951); R. Jost and W. Kohn, Phys. Rev. {\bf 87}, 977 (1952).
%
\bibitem{Calarco06}
Z. Idziaszek and T. Calarco, Phys. Rev. Letters {\bf 96}, 013201 (2006).
%
\bibitem{Yu84}
C. C. Yu and P. W. Anderson, Phys. Rev. B {\bf 29}, 6165 (1984).
%
\bibitem{Clogston62}
A. M. Clogston, Phys. Rev. {\bf 125}, 439 (1962).
%
\bibitem{Vladar86}
K. Vlad\'ar, G. T. Zim\'anyi, and A. Zawadowski, Phys. Rev. Letters {\bf 56}, 286 (1986).
%
\bibitem{Vladar88}
K. Vlad\'ar, A. Zawadowski, and G. T. Zim\'anyi, Phys. Rev. B {\bf 37}, 2001 (1988);
{\it ibid}\, 2015 (1988).
%
\bibitem{Arnold07}
M. Arnold, T. Langenbruch, and J. Kroha, Phys. Rev. Letters {\bf 99}, 186601 (2007).
%
\bibitem{Doniach74}
S. Doniach and E. H. Sondheimer, {\it Green's Functions for Solid State Physicists}
(Benjamin, London, 1974). 
%


\end{thebibliography}
\end{document}